\documentclass[12pt,a4paper]{article}

\usepackage[intlimits]{amsmath}
\usepackage{amsmath}
\usepackage{amsfonts}
\usepackage{amssymb}
\usepackage[dvips]{graphicx}

\begin{document}

\title{Superbosonization\footnote{Presented at the Conference on Random Matrix Theory: From Fundamental Physics to Applications, Krak\'ow, Poland, May 2-6, 2007}}
   
\author{Hans-J\"{u}rgen Sommers}
\maketitle
\begin{flushleft}
{\it  Fachbereich Physik, Universit\"{a}t Duisburg-Essen\\
47048 Duisburg, Germany}
\end{flushleft}
 
\begin{abstract}
We give a constructive proof for the superbosonization formula for
invariant random matrix ensembles, which is the supersymmetry analog
of the theory of Wishart matrices. Formulas are given for unitary,
orthogonal and symplectic symmetry, but worked out explicitly only
for the orthogonal case. The method promises to become a powerful
tool for investigating the universality of spectral correlation
functions for a broad class of random matrix ensembles of
non-Gaussian type.
\end{abstract}
Pacs: 0250.-r, 0540.-a  \vspace*{0.5cm}\\
\section{Introduction}
I will report on superbosonization, a technique  in random matrix
theory (RMT), the appellation  given by Efetov et al. \cite{Efetov}
which we want to make a rigorous tool. Recent mathematical and theoretical work
from SFB/TR12 is to be published \cite{ELSZ,BEKYZ}. The method is the
supersymmetry analog of the theory of Wishart matrices. Other
people have done related work connected with universality properties
of invariant ensembles \cite{LSSS,HW,Fyodorov,Guhr}. But nowhere
so far the precise integration domain for the supermatrices under
consideration have been specified. Here I want to present a
constructive proof of the superbosonization formula on
a level, which can be understood by a physicist.

One may generate correlations of random Green functions from some
Gaussian integrals over a set of $N$-dimensional vectors building a
$N\times p$ dimensional matrix $\Phi$, which after averaging are no
longer Gaussian, but the integrand  is a function of the  
matrix $Q = \frac{1}{N} \Phi^\dagger \Phi$ of invariants. The transition to this
new set of variables is called bosonization and it makes e.g.
large-$N$ evaluations simpler using saddle-point techniques.
This kind of order parameter was previously introduced in the theory of
spin glasses and the theory of Anderson localization. These theories
involve the socalled replica trick, where at the end of calculation,
after averaging over the ensemble, one has to take the limit of
number of replicas to zero, a procedure that is very difficult to
make
mathematically rigorous. 

A more rigorous method is to  introduce
supervectors $\Psi$ (if possible) with commuting and anticommuting
components. Then no replica limit is needed, in some sense the
dimension of
Grassmannian (anticommuting) vectors can be considered as negative.
The change of variables from a set of supervectors $\Psi$ to a
supermatrix $Q = \frac{1}{N} \Psi^\dagger \Psi$ is called
superbosonization. It is useful for invariant ensembles and
non-Gaussian ensembles and avoids the sometimes cumbersome
Hubbard-Stratonovic transformation.
\section{Superbosonization formula}
Let us assume, that we want to integrate a function $F$ of
invariants (with respect to a symmetry group)
\begin{equation}
\label{one}
 I_{pq} = \int D\zeta \, D\Phi \;
F\left(\begin{array}{cc}
\zeta^\dagger \zeta&\zeta^\dagger \Phi\\
\Phi^\dagger \zeta& \Phi^\dagger \Phi
\end{array}
\right)\, .
\end{equation}
Here $\zeta$ are anticommuting variables building a rectangular
$N\times p$ matrix ($p$ even), $\Phi$ are commuting variables
building a $N\times q$ matrix (and we need lateron  $N \geqslant
q$). Originally one starts with an invariant function of $\zeta$ and
$\Phi$ and ends up with a function of invariants, which is not
unique for Grassmannians, nevertheless the result of integration is
unique. $D\zeta$ is the Berezin integration form and $D\Phi$ is
the flat  measure of matrix elements.

We want to transform to integrations over supermatrices $Q=
\bigl(\begin{smallmatrix} A&\sigma^\dagger\\ \sigma&B
\end{smallmatrix}\bigr)
 $ with bosonic (commuting) entries
$A$, $B$ and fermionic (anticommuting) entries $\sigma$, $\sigma^\dagger$. In this
way we reduce
 considerably the number of bosonic and fermionic integration variables especially
 in the case where $N$ goes very large.
 It is not surprising that $B$ runs over positive Hermitian matrices and $\sigma$ and $\sigma^\dagger$
 run over Grassmannians. But the amazing thing is that the entry in the Fermi-Fermi sector,
 which is originally nilpotent, is replaced by a matrix A which runs
 over a manifold of unitary matrices. There is a compact way of
 writing the integral (\ref{one}) as superintegral containing some
 power  $M$ of the socalled superdeterminant  ${\rm S}\det Q$ of the
 supermatrix Q:
 \begin{eqnarray}
 \nonumber
I_{pq} &=& {\cal N} \int DA \int DB \int D(\sigma,  \sigma^\dagger)
\left[ \frac{\det B}{\det (A - \sigma^\dagger \frac{1}{B}
\sigma)}\right]^M F\left( \begin{array}{ll} A & \sigma^\dagger \\
\sigma& B
\end{array} \right) \\
&=& {\cal N} \int DQ\  ({\rm S}\det Q)^M\  F(Q)
 \end{eqnarray}
 The appearing measures are the flat measures and $\cal N$ is
a  normalization constant. This is the superbosonization formula and
I want to specify it for different symmetry groups.

For the real orthogonal group $O_N$ the dagger~$^\dagger$ just means
the transposed~$^T$, $A$ is skew-symmetric ($ A = - A^T$) and
unitary, and $B$ is real symmetric ($B = B^T$) and positive. The
power $M$ is given by $M = (N+p - q - 1)/2$.\\
For the unitary symplectic group $U \mbox{Sp}_{N}$  ($N$ even) the
dagger~$^\dagger$ means the dual~$^D$, with e.g. $A^D = Z A^T Z^T$
where $Z$ is the symplectic unit which is quasidiagonal with
quaternion elements 
$\bigl( \begin{smallmatrix}  0&1\\ -1&0 \end{smallmatrix}\bigr)$ on
the diagonal. $A$ is anti-selfdual ($A = -A^D$) and unitary and $B$
is Hermitian, selfdual ($B = B^D$) and positive. The power $M$ is
given by $M = (N + p -q +1)/2$. In the case of unitary group $U_N$
the dagger means just the usual adjoint = transposed and complex
conjugate. The Grassmannian $\sigma^\dagger$ can be any independent
set $\widetilde{\sigma}$ of Grassmannians. Here $M = N + p - q$. 

Let us say a bit on the manifolds. $A = - A^T$ implies that $(ZA)^D
= (ZA)$, i.e. for  orthogonal symmetry $ZA$ belongs to the set of
selfdual unitary matrices, which is called in physics the circular
symplectic ensemble CSE (not only meant as an invariant measure but
also as a manifold). Similarly $A = -A^D$ implies that $ZA$ is
symmetric. Thus for symplectic symmetry $ZA$ belongs  to the
circular orthogonal ensemble COE. Finally for unitary symmetry $ZA$
belongs to the circular unitary ensemble CUE.

On the other hand the matrices $B$ are positive Hermitian and
symmetric or selfdual in the orthogonal or symplectic case. All
those manifolds are Riemannian symmetric spaces, which were under
special consideration in our SFB/TR12, related to the socalled
ten-fold way of universalities in RMT for fermionic systems
\cite{Altland}. Moreover the supermatrices 
$\widetilde{Q} = 
\bigl( \begin{smallmatrix} Z&0 \\ 0& 1 \end{smallmatrix}\bigr) Q$
 belong to Riemannian symmetric superspaces
introduced by Zirnbauer
\cite{Zirnbauer}.

The measures which we have indicated so far are the flat ones and
they are related to the invariant measures on the corresponding
manifolds by some power of determinant or superdeterminant. E.g. the
invariant measures on the supermanifolds are (up to normalization)
\begin{equation}
d\mu(Q) = DA \, DB \,\  D(\sigma, \sigma^\dagger)\  ({\rm S}\det Q)^R
\end{equation}
with $R = p-q$ for $\beta =2 $, $R = (p - q - 1)/2$   for $\beta =1
$, $R = (p - q + 1)/2$  for $\beta = 4$. The invariant measures on
the manifolds of $A$ and $B$ can be read off
from these expressions by putting $q =0 $ or $p = 0$.

Let us say a little bit about the normalization constant
\begin{equation}
{\cal N} = C(N,q) \cdot D(N-q,p)
\end{equation}
It factorizes in two constants coming from the Bose-Bose sector
$C(N,q)$ and from the Fermi-Fermi sector $D(N-q,p)$. They are
essentially ratios of volumes of the corresponding symmetry groups.
They are separately only defined for $N\geqslant q$ because $N-q$ is
the smallest dimension of symmetry group that appears. And the proof
shows that this is actually needed. However, the product ${\cal N} =
C \cdot D$  is even defined for $N + p - q \geqslant 0$, it is not
clear if this has physical relevance.
\section{Idea of proof}
Let me now give an idea of the proof of the superbosonization
formula. We start with $p=0$, no Grassmannians, this is just the
case of Wishart matrices. In this case one has to integrate some
function $H(\Phi^\dagger \Phi)$ over bosonic vectors $\Phi$:
\begin{eqnarray}
\nonumber \label{five}
 \int D \Phi H(\Phi^\dagger \Phi) &=& \int DB D\Phi \;
\delta(B -
\Phi^\dagger \Phi) H(B)\\
&=& C(N,q) \int DB \, H(B) (\det B)^M
\end{eqnarray}
The last equation can simply be found by rescaling $\Phi = \Phi'
\sqrt{B}$ with $B > 0$ for $N \geqslant q$. Since $C(N,q)$ is
independent of the function $H(B)$ we may choose $H(B) = \exp( -
\mbox{Tr } B)$ and find for example with $\beta =1 $ using Selberg's
integral \cite{Mehta} and diagonalization  of $B$ with orthogonal
matrices:
\begin{eqnarray}
\nonumber \frac{\pi^{N q/2}}{C(N,q)} &=& \int\limits_{B>0} DB \,
{\rm
e}^{-{\rm Tr } B} \det B^{(N-q-1)/2}\\
 &=& \frac{1}{2^{q/2}}
\prod_{j =0}^{q-1} \pi^{j/2} \Gamma((N-q+1 +j)/2) \, .
\end{eqnarray}
This constant  is related to the ratio of volumes of orthogonal
groups\\ $V(O_{N-q})/V(O_N)$. Since the manifold $B>0$ is noncompact
there appear noncompact integrals over eigenvalues of $B$: $b_i >
0$.

More interesting is the case $q = 0$, only Grassmannians, the case
of Grassmann Wishart matrices $\zeta$: $(N-q) \times p$, $p$ even.
In this case the naive introduction of a $\delta$-function for $A =
\zeta^T \zeta$ leads to inconsistencies and we have to be more
careful. We have to integrate a function $G(\zeta^T \zeta)$ which we
write with the help of a shift operator
\begin{eqnarray}
\nonumber \int  D \zeta \, G(\zeta^T \zeta) & =& \int D \zeta \left.
\exp \left( \mbox{Tr } \zeta^T \zeta
\frac{\delta}{\delta A} \right) G(A) \right|_{A=0}\\
\label{pfaff} &=& \left. \sqrt{\det \left( 2\frac{\delta}{\delta A}
\right)}^{N-q} G(A)\right|_{A=0}\ .
\end{eqnarray}
Here $A$ is an antisymmetric matrix and the $\zeta$-integration
yields the $(N-q)$-th power of the Pfaffian $\sqrt{\det \left(
2\frac{\delta}{\delta A} \right)}$.

The superbosonization formula to be proven says:
\begin{equation}
\label{eight}
 \int D\zeta \, G(\zeta^T \zeta) = D(N-q,p) \int
\frac{DA}{(\det A)^{\frac{N-q+p-1}{2}}} G(A)
\end{equation}
where $A$ runs over the unitary antisymmetric matrices. $G(A)$ may
be any polynomial or analytic function. Since $A$ is unitary there
is no problem at $A=0$ or $\det A = 0$ for this integral. Obviously
it is enough to prove this formula for any exponential function
$G(A) = \exp (\mbox{Tr } AB/2)$ where $B$ is again antisymmetric. In
this case the Pfaffian action (\ref{pfaff}) can simply be calculated
and the last equation can again be proven by rescaling choosing $ZB$
a selfdual unitary matrix. To calculate the constant $D(N-q, p)$ one
may choose $B = Z$. Then one finds diagonalizing  $ZA$ with
symplectic matrices
\begin{eqnarray}
\nonumber \frac{(-1)^{(N-q)p/2}}{D(N - q,p)} &=& \frac{V(U
\mbox{Sp}_p)}{V(U \mbox{Sp}_2)^{p/2} \bigl( \frac{p}{2} \bigr)!}
\cdot \oint \prod_{i<j} (a_i - a_j)^4 \prod_{k=1}^{p/2} da_k \,
a_k^{-(N-q+p-1)} {\rm e}^{a_k}\\
&=& \left( \frac{2 \pi i}{2^{N-q}} \right)^{p/2} \prod_{j=0}^{p-1}
\frac{\pi^{j/2}}{\Gamma ((N-q+1 + j)/2)} \, .
\end{eqnarray}
This is again related to the ratio of volumes of real orthogonal
groups\\ $V(O_{N-q+p})/V(O_{N-q})$. Since the
manifold $ZA \in CSE$ is compact there appear integrals over
eigenvalues $a_i$ of $ZA$ along the unit circle. Interestingly,
these compact integrals in the complex plane are related to the
noncompact
integrals with $b_i > 0$ along the real axis.

Finally let me consider the full supermatrix $Q$. I only report here
on the orthogonal case. In the integral
\begin{equation}
I_{pq} = \int D\zeta \int D\Phi  \; F\left( \begin{array}{cc}
\zeta^T \zeta& \zeta^T \Phi\\
\Phi^T \zeta& \Phi^T \Phi\end{array} \right)
\end{equation}
we have to go from $N\times (p,q)$ variables $(\zeta, \Phi)$ to the
$(p,q) \times (p,q)$ supermatrix $Q$. We choose an orthogonal matrix
$O$ which rotates $\Phi$ to a quadratic $q\times q$ matrix
$\sqrt{B}$
\begin{equation}
\Phi = O \Phi_0 \,;  \; \Phi_0 = \left( \begin{array}{c} 0\\
\sqrt{B}
\end{array} \right) \, .
\end{equation}
Then the Grassmannians are rotated correspondingly
\begin{equation}
O^T \zeta = \widetilde{\zeta} = \left(\begin{array}{c} \zeta_1\\
\zeta_0 \end{array} \right)
\end{equation}
which has a $q\times p$ component $\zeta_0$ which transforms to
$\sigma$
\begin{equation}
\sigma = \Phi^T \zeta = \Phi_0^T \widetilde{\zeta} = \sqrt{B}
\zeta_0
\end{equation}
and a remaining $(N-q) \times p$ component $\zeta_1$ which can be
integrated out using (\ref{eight}) and
\begin{equation}
\zeta^T \zeta = \widetilde{\zeta}^T \widetilde{\zeta} = \zeta^T_1
\zeta_1 + \zeta_0^T \zeta_0 \, .
\end{equation}
Similarly the integration over $B$ follows from (\ref{five}). The
result is almost what we want
\begin{equation}
I_{pq} = {\cal N} \int DA \int DB \int D\sigma \left[\frac{\det B}{\det A}
\right]^{(N - q + p -1)/2} F \left( \begin{array}{cc} A + \sigma^T
\frac{1}{B} \sigma& \sigma^T\\
\sigma& B \end{array} \right)\, .
\end{equation}
Now shift $A \rightarrow A - \sigma^T \frac{1}{B} \sigma$, which is
possible since the manifold CSE has no boundary (compact symmetric
space) and we end up with
\begin{equation}
I_{pq} ={\cal N}  \int DQ \ ({\rm S}\det Q)^{(N-q + p - 1)/2}\  F(Q)
\end{equation}
with ${\rm S}\det Q = \det B/ \det(A - \sigma^T \frac{1}{B} \sigma)$. This
completes the proof for the orthogonal case. Similar considerations
lead to the corresponding results for the symplectic and unitary
cases.


\begin{thebibliography}{10}
\bibitem{Efetov} K.B. Efetov, G. Schwiete and  K. Takahashi, Phys. Rev.
Lett {\bf 92} 026807 (2004)
\bibitem{ELSZ}   P. Littelmann, H.-J.Sommers and M.R. Zirnbauer, arXiv:0707.2929v1[math-ph]
\bibitem{BEKYZ}      J.E. Bunder, K.B. Efetov, V.E. Kratvtsov, O.M. Yevtushenko, and M.R. Zirnbauer, arXiv:0707.2932v1[cond-mat.mes-hall]
\bibitem{LSSS} N.Lehmann, D. Saher, V.V. Sokolov, H.-J. Sommers, Nucl. Phys. A {\bf 582} 223 (1995) 
\bibitem{HW} G. Hackenbroich and H.A. Weidenm\"uller, Phys. Rev.
Lett {\bf 74} 4418 (1995)
\bibitem{Fyodorov} Y.V. Fyodorov, Nucl. Phys. B{\bf 621} 643 (2002)
\bibitem{Guhr} T. Guhr, J. Phys. A{\bf 39} 13191 (2006)
\bibitem{Altland} A. Altland and M.R. Zirnbauer, Phys. Rev. Lett.
{\bf 76}, 3420 (1996)
\bibitem{Zirnbauer} M.R. Zirnbauer, J. Math. Phys {\bf 37} 4986
(1996)
\bibitem{Mehta} M.L. Mehta, Random Matrices, Academic Press, Inc
(1991)
\end{thebibliography}
\end{document}